\documentclass[aip,prl,preprint,superscriptaddress]{revtex4}
\usepackage{graphicx}

\begin{document}

\title{One-dimensional spin texture of Bi(441); Quantum Spin Hall properties without a topological insulator}

\author{M.~Bianchi}
\affiliation{Department of Physics and Astronomy, Interdisciplinary Nanoscience Center (iNANO), University of Aarhus, 8000 Aarhus C, Denmark}
\author{F.~Song}
\affiliation{Zernike Institute of Advanced Materials, University of Groningen, 9747 AG, The Netherlands.}
\affiliation{Shanghai Institute of Applied Physics, Chinese Academy of Science, 201204, P. R. China}
\author{S. Cooil}
\affiliation{Department of Physics, Aberystwyth University, Aberystwyth, UK. SY23 3BZ}
\author{\AA.F. Monsen}
\affiliation{Dept. of Physics, Norwegian University of Science and Technology (NTNU), Trondheim, Norway.}
\author{E.~Wahlstr\"om}
\affiliation{Dept. of Physics, Norwegian University of Science and Technology (NTNU), Trondheim, Norway.}
\author{J.A.~Miwa}
\affiliation{Department of Physics and Astronomy, Interdisciplinary Nanoscience Center (iNANO), University of Aarhus, 8000 Aarhus C, Denmark}
\author{E.D.L.~Rienks}
\address{Helmholtz-Zentrum Berlin, Albert-Einstein-Str.\ 15, 12489 Berlin, Germany}
\author{D.A.~Evans}
\affiliation{Department of Physics, Aberystwyth University, Aberystwyth, UK. SY23 3BZ}
\author{A.~Strozecka}
\affiliation{Institut f\"ur Experimentalphysik, Freie Universit\"at Berlin, Arnimallee 14, 14195 Berlin, Germany}
\author{J.I.~Pascual}
\affiliation{CIC nanoGUNE, 20018 Donostia-San Sebasti\'an, Spain}
\author{M.~Leandersson}
\affiliation{MAX IV Laboratory, Lund University, P.O. Box 118, 221 00 Lund, Sweden}
\author{T.~Balasubramanian}
\affiliation{MAX IV Laboratory, Lund University, P.O. Box 118, 221 00 Lund, Sweden}
\author{Ph.~Hofmann}
\affiliation{Department of Physics and Astronomy, Interdisciplinary Nanoscience Center (iNANO), University of Aarhus, 8000 Aarhus C, Denmark}
\author{J.W.~Wells}
\email[]{quantum.wells@gmail.com}
\affiliation{Dept. of Physics, Norwegian University of Science and Technology (NTNU), Trondheim, Norway.}

\begin{abstract}
The high index (441) surface of bismuth has been studied using Scanning Tunnelling Microscopy (STM), Angle Resolved Photoemission Spectroscopy (APRES) and spin-resolved ARPES. The surface is strongly corrugated, exposing a regular array of (110)-like terraces. Two surface localised states are observed, both of which are linearly dispersing in one in-plane direction ($k_x$), and dispersionless in the orthogonal in-plane direction ($k_y$), and both of which have a Dirac-like crossing at $k_x$=0. Spin ARPES reveals a strong in-plane polarisation, consistent with Rashba-like spin-orbit coupling. One state has a strong out-of-plane spin component, which matches with the miscut angle, suggesting its {possible} origin as an edge-state. The electronic structure of Bi(441) has significant similarities with topological insulator surface states and is expected to support one dimensional Quantum Spin Hall-like coupled spin-charge transport properties with inhibited backscattering, without requiring a topological insulator bulk.
\end{abstract}

\maketitle

Bismuth is an interesting material for many reasons \cite{Edelman:1996,Yang:1999a,Rogacheva:2003,Hofmann:2006}. Being the heaviest stable element, spin-orbit interactions are especially significant \cite{Gonze:1988}, making it an ideal platform for testing fundamental concepts, such as spin non-degeneracy limiting impurity scattering \cite{Strozecka:2011} and inhibiting a charge density wave \cite{Kim:2005b}. The heavy nuclear weight can also cause a parity inversion such that semiconducting bismuth alloys can support an inverted band gap and the creation of a topological insulator phase \cite{Zhang:2009}. Such topological insulators have attracted much interest \cite{Moore:2010} and have been suggested as candidates for a range of potential device applications \cite{Cha:2010,Xiu:2011}. 

Bismuth and many of its compounds are layered materials. As with graphene, ultra thin bismuth is reported to be a two-dimensional (2D) topological insulator \cite{Qiao:2011a,Wada:2011,Murakami:2006,Chen:2013}, supporting an edge-localised Quantum Spin Hall (QSH) like state.   The vicinal surface of bismuth, Bi(114) \cite{Wells:2009,Leuenberger:2013a}, consists of an array of such edges, and supports a single one-dimensional Fermi-contour, which has been revealed to be non-degenerate with respect to spin and is reminiscent of a topological insulator state. 

\begin{figure}
\includegraphics[width=0.75\columnwidth]{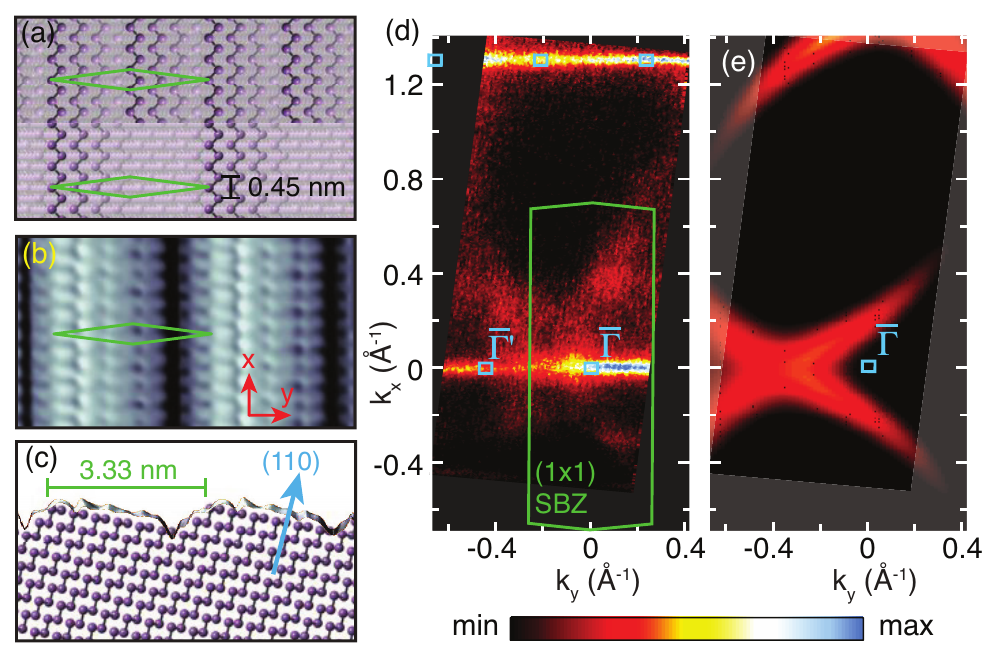}
\caption{(a) Truncated bulk model of the surface, showing the $1\times1$ (upper half) and the missing row (lower half) reconstruction. The $1\times1$ unit cell is superimposed in green. (b) STM image (bias voltage = 0.091~V, tunnelling current = 90~pA.), similarly scaled with the same unit cell overlaid. (c) side-view of the $2\times2$ truncated bulk model and the measured STM height profile. (d) Fermi surface map performed with photon energy $h\nu=70$~eV, showing the $1\times1$ SBZ, and the $\overline\Gamma$ points. Bright colours (light blue, white, yellow) indicate high intensity and dark colours (black, dark red) indicate low intensity. (e) Tight binding calculated constant energy surface, indicating where the projected bulk states are close enough to the measured Fermi surface for a weak intensity contribution to be expected (using the parameters of Liu and Allen \cite{Liu:1995}).}
\label{realspace}
\end{figure}

Here, we investigate the surface electronic structure and spin texture of another vicinal bismuth surface; Bi(441). A scanning tunnelling microscopy (STM) study was performed, and the rhombic surface unit cell of Bi(441) is seen. The (441) surface is observed to support a $2\times2$ reconstruction which is well described by a simple truncated bulk, `missing-row' model (see Fig.~\ref{realspace}(a-c)), thus the surface can be viewed as a regular 1D array of (110)  oriented domains, with edges of the (111) bilayers exposed.

Using angle resolved photoemission spectroscopy (ARPES) we show that the surface electronic structure is strongly influenced by the vicinality of the surface. In fact, the Fermi surface consists almost entirely of a single 1D feature which passes through the surface brillouin zone (SBZ) centre, $\overline{\Gamma}$ (see Fig.~\ref{realspace}(d)). As in the case of Bi(114), the Fermi surface shows no sign of the $2\times2$ surface reconstruction. Whilst the periodicity in $k_x$ is clear, there is a lack of any observable dispersion in $k_y$. The lack of periodicity in the surface Fermi surface highlights the model 1D behaviour of the sample since $k_y$ becomes an irrelevant parameter in the perfect 1D case. 

An additional weak and broad feature (visible as a faint `X') is also present in the ARPES measurement (Fig.~\ref{realspace}(d)). By considering its dispersion with photon energy, it can be attributed to a projected bulk state. This view is strongly supported by tight binding calculations (see Fig.\ \ref{realspace} (e), and the Supplementary Material).  There is a lack of bulk symmetry in the $y$-direction which hinders a simple identification of $k_y=0$. However, the tight binding calculated bulk band projection allows $\overline{\Gamma}$ to be identified. 

The single 1D feature in the Fermi surface raises the question of the nature of this state. Strong spin-orbit coupling is expected \cite{Hofmann:2006}, and states reminiscent of topological insulator surface have been observed on other bismuth \cite{Wells:2009,Ast:2001,Agergaard:2001} (and Bi$_x$Sb$_{1-x}$ \cite{Agergaard:2001,Zhu:2013}) surfaces. However, {if bulk bismuth is described as topologically trivial \cite{Hofmann:2005,Teo:2008,Ohtsubo:2013} then an unpaired Fermi level crossing is not expected here}.

Since the surface can be viewed as an array of (111) bilayer edges, an alternative approach is to consider the arguments of topology in a similar way; bilayer bismuth (111) is described as topologically non-trivial in its 2D bulk, thus its 1D edges can be expected to support a topological state. In a simplified view, the high index surfaces of the form $(x,x,1)$ behave as an array of such (111) bilayer edges, coupled only by van der Waals forces. The lack of strong interaction between the layers is manifest in the 1D Fermi surface, thus it is perhaps reasonable to expect that the 1D state inherits the characteristics of the bilayer's topological edge state \cite{Murakami:2006,Drozdov:2014}.

\begin{figure}[b]
\includegraphics[width=0.75\columnwidth]{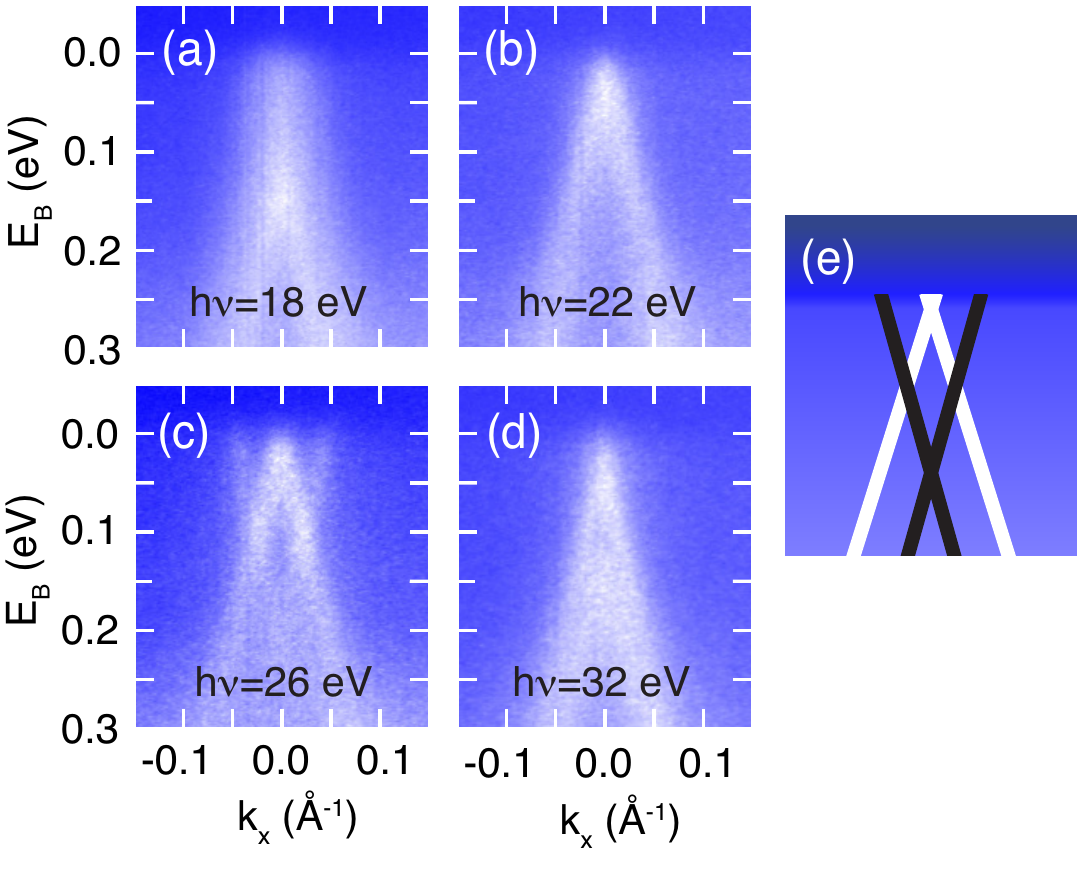}
\caption{(a), (b), (c) and (d) ARPES measurements performed at $k_y$=0 \AA$^{-1}$ and photon energies 18, 22, 26 and 32 eV, respectively. (e) Schematic depicting the presence of `inner' (white) and `outer' (black) Dirac-like states.}
\label{arpes_ph}
\end{figure}

Measurements made at a range of photon energies (Fig.\ \ref{arpes_ph}) reveal the existence two Dirac-like states, with very similar dispersions in the $k_x$ direction. Their cross-sections vary strongly, thus typically only one state dominates the measurement. At particular photon energies, both states are visible, for example at $h\nu=26$~eV (Fig. \ref{arpes_ph}(c)). A likely cause of this strong variation in intensity is a corresponding variation in the matrix element describing the photo-excitation to a bulk-like final state \cite{Miwa:2013}. The lack of dispersion with photon energy (or equivalently, $k_\perp$ \cite{Himpsel:1980}) confirms that these states are surface or edge states. Our tight binding calculations of the bulk states (which are otherwise in good agreement with the measurements) do not reproduce these states, further supporting the notion that neither of these states are bulk-like.

\begin{figure}
\includegraphics[width=0.75\columnwidth]{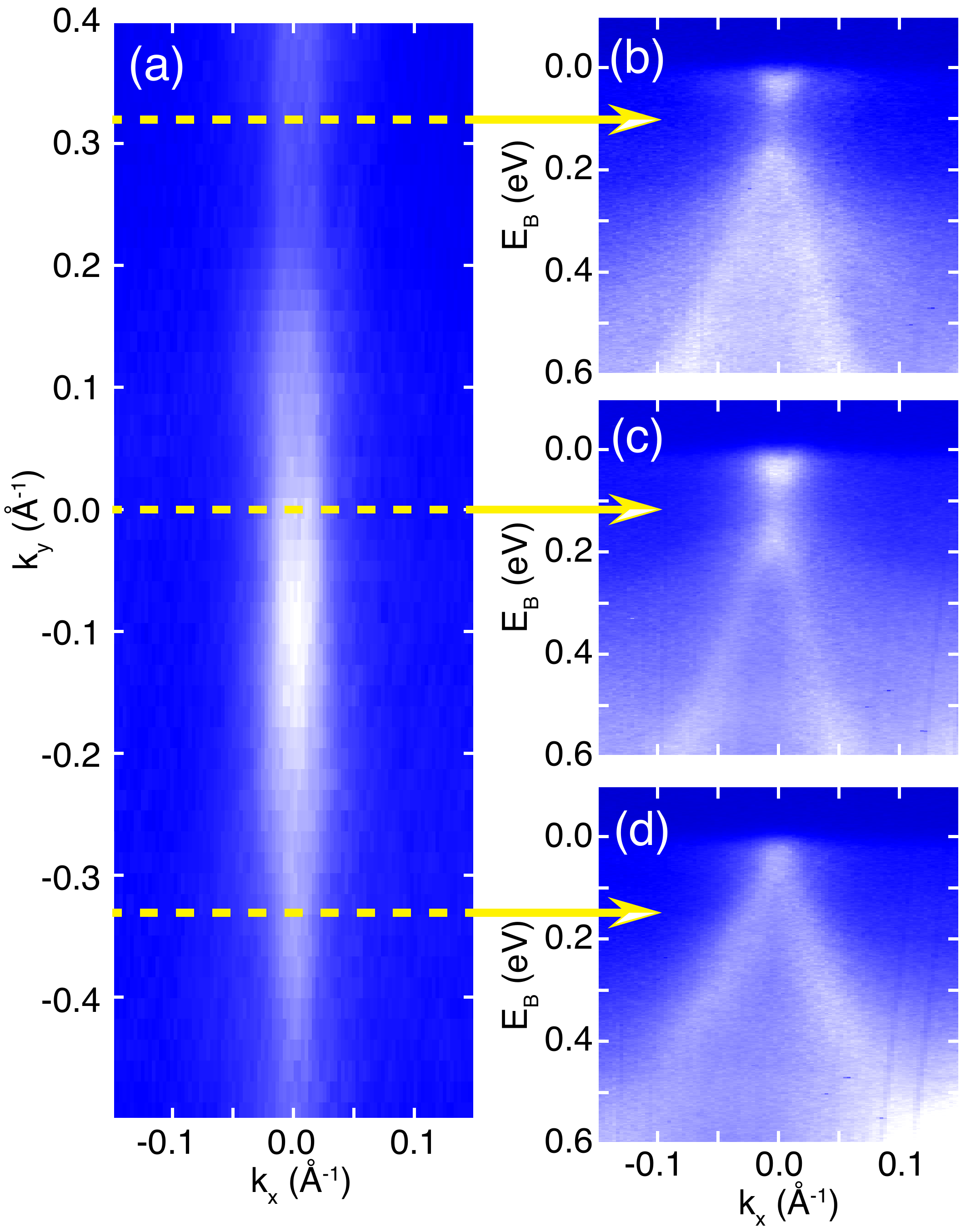}
\caption{(a) Fermi surface collected with photon energy 19~eV. Yellow horizontal lines indicate where the slices shown in (b), (c) and (d) have been extracted ($k_y$=+0.33, $k_y$=0 and $k_y$=-0.33 \AA$^{-1}$, respectively).}
\label{arpes_ky}
\end{figure}

\begin{figure}
\includegraphics[width=0.6\columnwidth]{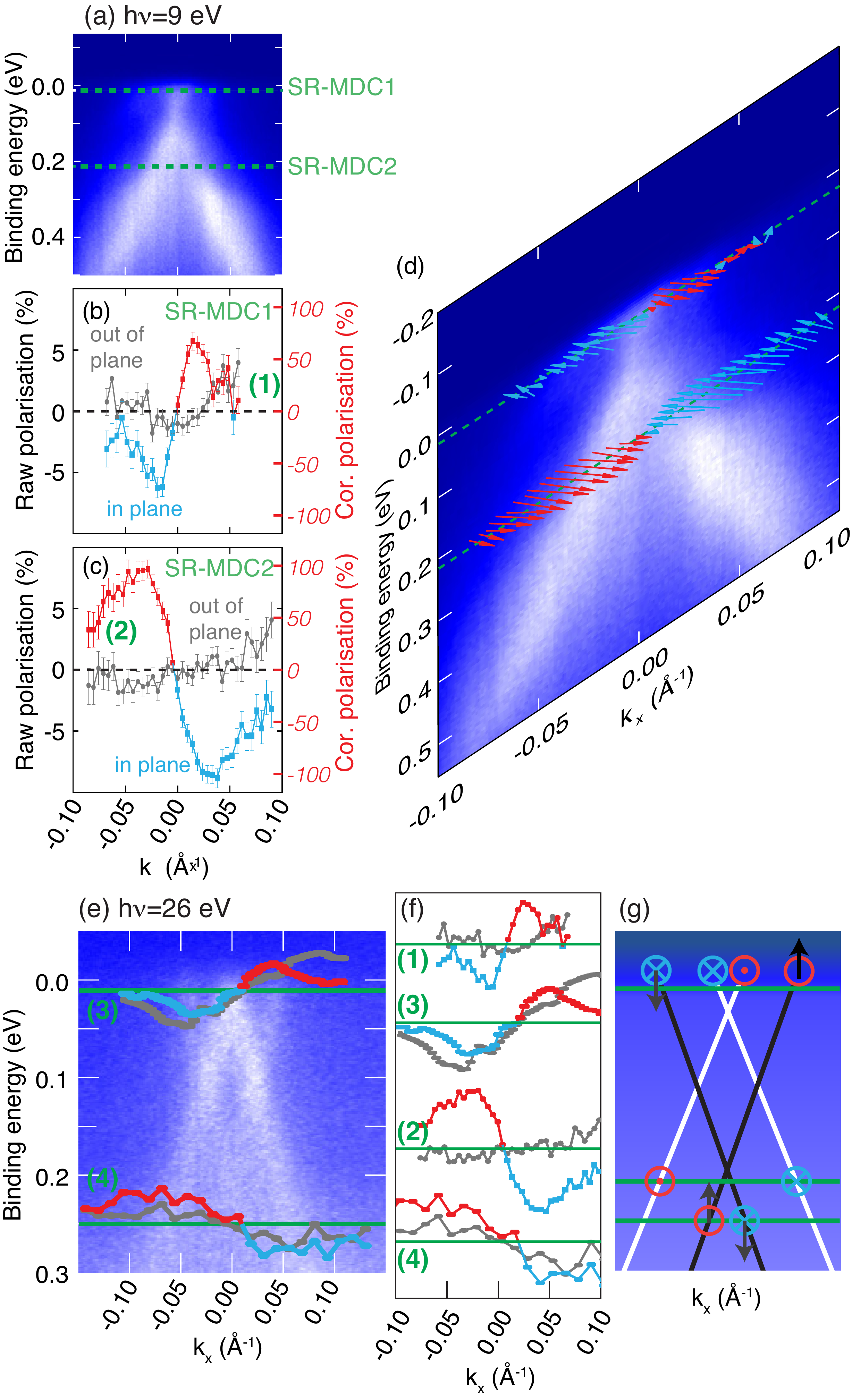}
\caption{(a) ARPES measurement performed at $h\nu=9$~eV showing where two Spin Resolved MDCs (SR-MDC1 and SR-MDC2) were collected (green dashed lines). (b) and (c) the raw spin polarisations collected at the Fermi level, and at $E_b=0.2$~eV, respectively, showing the in-plane (blue-red) and out-of-plane (grey) spin components. {The red axis indicates \textit{corrected polarisation},  found by subtracting a non-polarised background and correcting for the Sherman function of 17~$\%$}. (d) ARPES measurement with the reconstructed spin vectors superposed. (e) In-plane (blue-red) and out-of-plane (grey) spin components from SR-MDCs atop an ARPES measurement at $h\nu=26$~eV, where the contribution of the outer state is stronger. In both cases $k_y$=0.  (f) Similarly scaled polarisation curves for the four SR-MDCs numbered in (a-c) and (e).  (g) Schematic representation of the measured spin vectors, and the positions at which they are measured.}
\label{spin_comb}
\end{figure}

The photoemission intensity at the Fermi level (Fig.~\ref{arpes_ky}(a)), collected at a photon energy of 19~eV, shows that this state has a weak dependence on $k_y$. This is best seen in constant $k_y$ slices taken at $k_y$=+0.33, $k_y$=0 and $k_y$=-0.33 \AA$^{-1}$, which reveal a pair of linearly dispersing branches (see Fig. \ref{arpes_ky}(b-d)), with Dirac-like crossings at $k_x$=0 with $E_B$$\approx$150, $E_B$$\approx$150 and $E_B$$\approx$0~meV for $k_y$=+0.33, $k_y$=0 and $k_y$=-0.33 \AA$^{-1}$, respectively. 

In fact, over the range of $k_y$ probed (Fig. \ref{arpes_ky}(a)), only a very weak dispersion is seen. Instead the intensity of the state at deeper binding energy,  and the state with the crossing nearest the Fermi level, depends  on $k_y$. Thus, it appears that both states have a true 1D nature, and very similar linear dispersions in $k_x$. The band structure of both states is presented schematically in Fig.\ \ref{arpes_ph}(e).

Since strong spin-orbit coupling is expected in this material \cite{Wells:2009,Hofmann:2006,Kim:2005b,Koroteev:2004}, we probe the spin polarisation of the states observed here by Spin-ARPES, using a Mott-type detector \cite{Berntsen:2010}. The small $k_x$ separation of the observed states requires good experimental resolutions, with measurements at low photon energy being preferable. For example, at photon energy 9~eV, the spin polarisation can be efficiently measured -- the state with its crossing point closest to the Fermi level (marked in white in Fig \ref{arpes_ph}(e)) is most intense, and hence dominates the ARPES measurement, although a very weak remnant of the outermost state is barely visible (see Fig. \ref{spin_comb}(a)). Correspondingly, this state dominates the spin-ARPES measurement, and hence its polarisation can be found unambiguously. The spin-ARPES measurements were performed as two momentum distribution curves (MDCs), one at the Fermi level, and the other around 200~meV below. A 2D Mott detector is oriented such that the out-of-plane spin component, and the in-plane component in the $y$-direction are measured. The raw spin-polarisation $P$ is given by $P=\chi (N_{+}-N_{-})/(N_{+}+N_{-})$, where $N_{+}$ and $N_{-}$ are the number of counts recorded by the detectors in the relevant positive and negative directions (i.e. `up' and `down' for the in-plane polarisation and `left' and `right' for the out-of-plane polarisation). Since the four detectors are not equally sensitive, a sensitivity factor $\chi$ is found by evaluating $\chi=(N_{+}+N_{-})/(N_{+}-N_{-})$ well above the Fermi level. In other words, the polarisation of the state is evaluated relative to a background measurement. It should be noted that no other corrections are made; no Sherman function is assumed, hence `raw' polarisation is stated. 

At photon energy 9~eV (Fig. \ref{spin_comb}(a)), two MDCs are selected to cross the state above and below its degenerate crossing at $k_x=0$. In both spin MDCs, the maximum polarisation is seen at $k_x\approx\pm0.02$~\AA$^{-1}$ (Fig. \ref{spin_comb}(b) and (c)), corresponding to the maximum intensity of the state. In both cases, the out-of-plane component (grey) shows no significant polarisation, whereas the in-plane component (blue-red) shows a strong sign reversal; negative to positive near the Fermi level, and positive to negative at $E_b\approx0.2$~eV, typical of a Rashba-like coupling. The raw spin polarisation is between 6 and 8~\%, which, after allowing for a non-polarised background, and correcting for the Sherman function ($\approx 17~\%$) indicates that the state is very strongly polarised. 

In order to observe the spin-polarisation of the second state (with its crossing point at around $E_B$=200~meV and schematically depicted in black in Fig. \ref{arpes_ph}(e)), the observed strong intensity variation with photon energy is exploited. At $h\nu=26$~eV, this state appears more intense, and once again spin MDCs can be performed at  the Fermi level, and at higher binding energy ($\approx0.25$~eV). The maximum observed polarisation at the Fermi level is now seen at $k_x\approx0.05$~\AA$^{-1}$, corresponding to the maximum intensity of the outer state (Fig. \ref{spin_comb}(e)). As before, a strong reversal in the in-plane polarisation (from negative to positive at the Fermi level) is seen, indicating that the in-plane polarisation of both states is the same. Contrary to the previous case, the out-of-plane component is now of a similar magnitude to the in-plane component, indicating that the spin vector is approximately 45$^{\circ}$ to the sample surface. {The MDC at higher binding energy shows a weaker polarisation reversal, probably because of the proximity of a bulk state contributing unpolarised intensity to the background. However, this MDC shows a clear reversal of both the in-plane and out-of-plane components; from positive (at negative $k_x$) to negative (at positive $k_x$).}

The results from the spin MDCs are summarised in Fig. \ref{spin_comb}(f). The measured polarisations are superposed on the ARPES data for the two relevant photon energies, as well as being represented schematically. At the Fermi level, all states have a simple Rashba-like in-plane component, whereas the outer branches also have a large out-of-plane component. 

The picture of the electronic structure presented here is broadly consistent with other 1D systems in which SOC is important \cite{Barke:2006a,Tegenkamp:2012} . For example, previous studies of 1D metallic structures on high index silicon surfaces show Rashba-like coupling and a 1D Fermi contour \cite{Barke:2006a}.  However, these metallic  structures on high index silicon surfaces also show a coupling of the 1D states which is not seen for Bi(441), despite the separation of the 1D chains being significantly less in the present case \cite{Barke:2007}.  In our case of Bi(441), the Fermi contour shows ideal non-coupled 1D behaviour.

In the present case of Bi(441), the observed surface states have a dispersion which is more akin to  topological insulator surface states than the Rashba coupled free-electron like parabola of decorated vicinal silicon \cite{Barke:2006a,Tegenkamp:2012}. Since alloying bismuth, for example forming Bi$_{1-x}$Sb$_x$, can result in the formation of topological insulator phase \cite{Zhu:2014}, it is reasonable to assume that Bi$_{1-x}$Sb$_x$(441) could support an odd number of non-trivial states derived from the states observed here. 

{The implications of this surface electronic structure for the charge/spin transport properties are important. The spin texture of the bandstructure is such that a reversal of $k_x$ requires an accompanying spin reversal. i.e. $\epsilon(k_x,\uparrow)\Rightarrow\epsilon(-k_x,\downarrow)$. This means that unless there is a mechanism for exchanging spin angular momentum with an impurity, the probability of a charge carrier (with $E\approx E_F$ and $k_x \approx k_{x,F}$) being backscattered by an impurity approaches zero. Furthermore, there are no other surface states at the Fermi level which allow scattering process which conserve $|\mathbf{k}|$ and $E$, thus the probability of momentum conserving scattering events is expected to be vanishing small. Indeed, in our STM investigations, there are no visible indications of scattering around impurities. The lack of defect scattering, the spin-texture, and the 1D Fermi contour suggest that Bi(441) should have transport properties akin to quantum spin hall (QSH) materials \cite{Bernevig:2006b}, which are desirable for emergent spintronic applications \cite{Moore:2010,Pesin:2012}. } 


{Although the surface properties of Bi(441) show much promise for spintronics, there is a sufficient contribution of unpolarised bulk states to dominate the transport \cite{Wells:2008a}. In order to utilise the surface spin-transport properties in a real device, it would be necessary to reduce the contribution from the bulk. Two possibilities methods to achieve this are bulk alloying (for example with Sb, to open bulk band gap \cite{Nakamura:2011}) and by growth of a vicinal thin layer on a semiconducting substrate such as silicon \cite{Wells:2007,Ning:2014}.}




The existence of a finite out-of-plane spin component is also intriguing. In the simple topological insulator case, or indeed for simple Rashba-like SOC, the spin vector at the Fermi level is perpendicular to both the surface and the momentum vector, yielding an entirely in-plane spin vector. For the high index surface Bi(114), a significant out-of-plane component (30$^{\circ}$) has been reported \cite{Wells:2009}, which is comparable to the present case of around 45$^{\circ}$  for Bi(441). In both cases, this coincides with the miscut from the (111) plane. In other words, although superficially similar to the decorated vicinal silicon case \cite{Barke:2006a,Tegenkamp:2012}, the two states observed here are not simply SOC branches of the same state, but rather have quite separate origins; {it is tempting to speculate that the state with a strong out-of-plane component originates} from a topological-like edge state of a (111) layer \footnote{As with the Bi(111) edge state, the dispersion is approximately linear around the BZ centre, and the crossing point at $k_x=0$ is close to the Fermi level. However, the band dispersion $dE/dk_x$ observed here is significantly steeper \cite{Murakami:2006}}, and the state lacking an out-of-plane component to be the (441) surface state. 

In conclusion, we have presented the electronic structure of a high index surface of bismuth, with a strong 1D surface corrugation. Two 1D states are observed, both of which are non-degenerate with respect to their spin. Bulk bismuth {may not be} a topological insulator, but is nonetheless close to a topological-insulator transition, thus the states observed can be rationalised as topological-insulator-like, linearly dispersing, spin-polarised states and are expected to support QSH-like coupled spin-charge transport, desirable in emergent spintronic applications.\\

\textbf{Acknowledgements:} We thank N.A.~Vinogradov and A. B.~Preobrajenski for facilitating the motorisation of the SR-MDCs, and for helpful discussions. S.C.~acknowledges funding from MAX-IV laboratories and Ph.H.~acknowledges the VILLUM fonden for financial support. \\

\textbf{Methods:} Samples were prepared by mechanically cutting and polishing a bulk single crystal 45$^{\circ}$ from the (111) natural cleavage plane. Following mechanical polishing, the orientation was confirmed by Laue diffraction. The sample was then electropolished, introduced into the ultra-high vacuum chamber and cleaned by multiple cycles of Ar$^+$ sputtering and very gentle annealing until contaminants were beneath the detection threshold.

ARPES measurements were performed at beamline SGM3 of the synchrotron `ASTRID' in Aarhus, Denmark \cite{Hoffmann:2004}. Low photon energy ARPES and Spin-ARPES measurements were performed at beamline I3 of MAX-III  \cite{Berntsen:2010}. STM measurements were performed  at Freie Universit\"at Berlin using a custom-made instrument working in ultrahigh vacuum and at low temperature (5 K).  Tight Binding calculations were performed using the parameters of Liu and Allen \cite{Liu:1995}.

{Spin-ARPES were measured using linear-vertical polarised light which was incident upon the sample at 75$^{\circ}$ from the surface normal. In other words, the polarisation vector of the light is predominantly out-of-plane relative to the sample and with a small component in the $k_y$ direction. Measurements were also made using the same sample geometry and linear-horizontal light (i.e. in the in-plane $k_x$ direction) and the measured electron spin was the same within the uncertainty of the experiment.}


\end{document}